\documentstyle[prd,aps,eqsecnum,amsfonts]{revtex}
\begin{document}
\draft

\title{Quantum Scalar Field on
       the Massless (2+1)-Dimensional Black Hole Background.}
\author{Daniele Binosi}
\address{Dipartimento di Fisica, Trento, Italia}
\author{Valter Moretti \footnote{email: moretti@science.unitn.it}}
\address{Dipartimento di Matematica, Universit\`a di Trento \\ 
and Istituto Nazionale di Fisica Nucleare,\\ 
Gruppo Collegato di Trento, Italia}
\author{Luciano Vanzo  \footnote{email: vanzo@science.unitn.it}and
Sergio Zerbini \footnote{email: zerbini@science.unitn.it}}
\address{Dipartimento di Fisica, Universit\`a di Trento \\ 
and Istituto Nazionale di Fisica Nucleare,\\
Gruppo Collegato di Trento, Italia}
\date{September 14, 1998}
\maketitle

\begin{abstract}

The behavior of a quantum scalar field is studied in the metric 
ground state of the (2+1)-dimensional black hole of 
Ba\~nados, Teitelboim and Zanelli which contains a naked singularity. 
The one-loop BTZ partition function and 
the associate black hole effective entropy, the expectation value of 
the quantum fluctuation as well as the renormalized  expectation value of 
the stress tensor are explicitly computed in the framework of
the $\zeta$-function procedure.
This is done for all values  of the coupling with the curvature, 
the mass of the field and the temperature of the quantum state.
In the massless conformally coupled case, the found stress tensor 
is  used for determining the quantum back 
reaction on the metric due to the scalar field in the quantum 
vacuum state, by solving the semiclassical Einstein equations. 
It is finally argued that, within the framework of the
$1/N$ expansion, the Cosmic Censorship Hypothesis is implemented 
since the naked singularity  of the ground state metric is shielded by an 
event horizon created by the back reaction. 

\end{abstract}


\section{Introduction.}

Recently, the 3-dimensional gravity theory has been studied in detail. 
Despite the simplicity of the 3-dimensional case (absence of
propagating gravitons), it is a
common belief that it deserves attention  as a useful laboratory in order 
to understand several fundamental issues associated with the black hole 
entropy, such as its statistical origin and  horizon divergence problems 
(see, for example, \cite{beke,frol,thoof}). In fact, a black hole solution 
has been found by Ba\~nados, Teitelboim and
Zanelli \cite{bana92-69-1849}, the so called BTZ black hole; in
particular, 
the simple geometrical structure of this black hole solution, allows exact 
computations, since its Euclidean counterpart is locally isometric to the 
constant curvature 3-dimensional hyperbolic space ${\mathbb{H}}^3$. 
Furthermore, investigations in the 3-dimensional case seem to be relevant 
for several other  reasons, amongst which we would like to remind the
CFT/AdS 
correspondence \cite{maldawitt98}, and the fact that higher dimensional 
black holes can be related to the BTZ black hole (namely the near horizon 
geometry of these higher dimensional black holes is essentially the BTZ
one). 
With regard to this, the BTZ entropy issue has been recently 
reviewed in \cite{carlip98} (where a complete list of references can also
be 
found), and in \cite{lee98}; the quantum evolution of the BTZ black hole 
within a Kaluza-Klein reduction has instead been investigated in
\cite{odin98}.
 
In this paper  we shall discuss the behavior of a 
quantum scalar field propagating in the gravitational 
ground state of the  BTZ black hole 
(\emph{i.e.} the BTZ solution in the limit of a vanishing black hole
mass), 
generalizing to the non-conformally invariant case previous 
results 
obtained in
\cite{lifs94,stei94,shir94,mart97,mann97-55-3622,ichi95-447-340}.
We shall also
attempt to explore the possible relevance  of the quantum fluctuations
with 
regard to the issue of the cosmic censorship hipothesys,
since the BTZ ground state solution shows a naked singularity and,
presumably, it might be the final state at the end of the black-hole
evaporation process. 
It is worthwhile stressing that
the global topology of this ground state is completely different 
from the topology of the BTZ black hole, and thus
it could be dangerous,
in order to investigate the 
one-loop effective potential of a quantum scalar field in this background,
considering the results for a massive BTZ black hole
and take the limit $M \to 0$ naively;
as a consequence, we shall
compute all the quantities directly, employing the $\zeta$-function 
procedure.
This is true also for the expectation value of the stress tensor, since no 
good reasons were found for considering the  zero temperature thermal  
state as the only physically sensible one.\\                 
             
The content of the paper is organized as follows. 
In Sect.~II we briefly review the
geometry of the Euclidean BTZ black hole and its ground state. In
Sect.~III we 
present an elementary derivation of the heat-kernel and the
$\zeta$-function 
related to a Laplace-like operator necessary for the computation of the 
$\zeta$-function regularized functional determinant. 
In Sect.~IV, the one-loop relative partition 
function associated with the BTZ background and its ground state is 
computed and some comments on the effective black hole entropy are 
presented. In Sect.~V the computation of
the quadratic fluctuations of the scalar field is performed, and the 
expectation value of the associated stress tensor is evaluated in the
framework of the local $\zeta$-function approach. 
In Sect.~VI, the back reaction due to 
the quantum fluctuations is computed. The paper ends with
some concluding remarks in Sect.~VII, and with an appendix, where some 
computational technicalities are presented.

\section{The Euclidean BTZ Black Hole and its Ground State.}

Here, following \cite{carl95-51-622}, we summarize  the geometrical
aspects of
the non rotating BTZ black hole \cite{bana92-69-1849} and its
gravitational 
ground state, which are relevant for our discussion. \\ 
In the local coordinates $(t,r,\varphi)$, with 
\begin{eqnarray}
r\in(\sqrt{8GM}\ell,+\infty), \quad t\in(-\infty, +\infty),
\quad \varphi\in[0,2\pi), 
\end{eqnarray}
and $\varphi=0$ identified with $\varphi=2\pi$,
the static Lorentzian metric of the (non-rotating) BTZ black hole reads 
\begin{eqnarray}
ds_L^2=-\left( \frac{r^2}{\ell^2}-8GM \right) dt^2+\left(
\frac{r^2}{\ell^2}-8G M \right)^{-1}
dr^2+r^2 d\varphi^2
\:,\label{lm}\end{eqnarray}
where $M$ is the standard ADM mass and $\ell$ is a dimensional constant. 
Notice the couple of Killing fields $\partial_t$ and $\partial_{\varphi}$
which are respectively time-like and space-like.
A direct calculation shows that the metric above is a solution of the
3-dimensional vacuum Einstein's equations with  negative cosmological 
constant,\emph{i.e.}
\begin{eqnarray}
{\mathcal{R}}_{\mu \nu}=2\Lambda g_{\mu \nu}\,, 
\qquad {\mathcal{R}}=6\Lambda=-\frac{6}{\ell^2}.
\label{ee}
\end{eqnarray}
Thus, the sectional curvature $k$ is constant and negative, namely
$k=\Lambda=-1/\ell^2$. The metric~(\ref{lm}) has a horizon radius given by
\begin{eqnarray}
r_+= \sqrt {8GM}\ell,
\label{hr}
\end{eqnarray}
and it describes a space-time locally isometric to AdS$^3$.\\ 
An Euclidean section related to this choice of the coordinates 
is obtained by the Wick rotation $t \to i \tau$  ($\tau \in \mathbb{R}$)
and
reads
\begin{eqnarray}
ds^2=\left( \frac{r^2}{\ell^2}-\frac{r^2_+}{\ell^2}\right) d\tau^2+\left( 
\frac{r^2}{\ell^2}- 
\frac{r^2_+}{\ell^2} \right)^{-1}
dr^2+r^2 d\varphi^2.
\label{em}
\end{eqnarray}
Changing the coordinates $(\tau,r,\varphi)$ into  the $(y, x_1,x_2)$ ones,
by means of the transformation
\begin{eqnarray}
& & y=\frac{r_+}{r} e^{\frac{r_+\varphi}{\ell}}, \label{strange} \\ 
& & x_1+ix_2=\frac{1}{r} \sqrt{ r^2-r_+^2}
\exp\left(i\frac{r_+\tau}{\ell^2}
+\frac{r_+\varphi}{\ell}\right),
\label{cco}
\end{eqnarray}
the metric becomes that of the upper-half space representation of
${\mathbb{H}}^3$, \emph{i.e.}
\begin{eqnarray}
ds^2=\frac{\ell^2}{y^2}\left( dy^2+dx_1^2+dx_2^2 \right).
\label{hm}
\end{eqnarray}
Anyhow, the range of the coordinates is not the maximal one for 
${\mathbb{H}}^3$, since $y$ is bounded above because of the upper bound 
$\varphi<2\pi$ and the lower bound $r>r_+$. Nevertheless, we
can maximally extend the range of  the new coordinates
into $x_1,x_2 \in \mathbb{R}$ and $y\in \mathbb{R}_+$ obtaining the whole 
hyperbolic three space.
As a consequence, it is now obvious that, barring the identification 
$0\sim 2\pi$ in $\varphi$, the Euclidean section (\ref{em}) describes 
a manifold isometric to a sub-manifold of the hyperbolic space 
${\mathbb{H}}^3$. Actually we can say much more employing the theory of 
Lie's groups of isometries.\\ 
Recalling that the group of isometries of ${\mathbb{H}}^3$ is 
$SL(2,{\mathbb{C}})$, we shall consider a discrete subgroup 
$\Gamma\subset  PSL(2,{\mathbb{C}})\equiv SL(2,{\mathbb{C}})/
\{\pm {\textrm{Id}}\}$
(${\textrm{Id}}$ is the identity element), which acts discontinuously at
the point $z$
belonging to the extended complex plane ${\mathbb{C}}\bigcup\{\infty\}$. 
We also recall that a transformation $\gamma\in\Gamma$,
with $\gamma\neq {\textrm{Id}}$ and
\begin{eqnarray}
\gamma z=\frac{az+b}{cz+d}, \qquad ad-bc=1, \qquad a,b,c,d\in{\mathbb{C}},
\end{eqnarray}
is called elliptic if $(\mathop{\rm Tr}\nolimits\gamma)^2=(a+d)^2$
satisfies 
$0\leq(\mathop{\rm Tr}\nolimits\gamma)^2<4$, hyperbolic if 
$(\mathop{\rm Tr}\nolimits\gamma)^2>4$, parabolic if 
$(\mathop{\rm Tr}\nolimits\gamma)^2=4$ and loxodromic if 
$(\mathop{\rm Tr}\nolimits\gamma)^2\in{\mathbb{C}}/\left[0,4\right]$.
The element $\gamma\in SL(2,{\mathbb{C}})$ acts on ${\mathbf{x}}=(y,w)\in 
{\mathbb{H}}^3$, with $w=x_1+ix_2$, by means of the following
linear-fractional 
transformation
\begin{eqnarray}
\gamma{\mathbf{x}}=\left(\frac{y}{|cw+d|^2+|c|^2y^2},\frac{(aw+b)
(\bar c \bar w+\bar d)+a \bar c y^2}{|cw+d|^2+|c|^2y^2} \right).
\label{SS3}
\end{eqnarray}
The periodicity of the  angular coordinate $\varphi$   
in (\ref{strange}), which corresponds to a one-parameter group of
isometries,
allows one to describe the BTZ black hole manifold (\ref{em}) as the
quotient
${\mathcal{H}}^3\equiv{\mathbb{H}}^3/\Gamma$, $\Gamma$ being a discrete
group 
of isometry possessing a primitive element $\gamma_h \in \Gamma$ defined
by 
the identification
\begin{eqnarray}
\gamma_h(y,w)=(e^{\frac{2 \pi r_+}{\ell}} y,e^{\frac{2 \pi r_+}{\ell}}w)
\sim (y,w),
\label{yid}
\end{eqnarray}
induced by $0 \sim 2\pi$ in (\ref{strange}).
According to (\ref{SS3}), this corresponds to the matrix
\begin{eqnarray}
\gamma_h= \left(
\begin{array}{cc}
e^{\frac{\pi r_+}{\ell}} & 0 \\ 
0 & e^{-\frac{\pi r_+}{\ell}}
\end{array}
\right),
\label{p}
\end{eqnarray}
namely to a hyperbolic element consisting in a pure dilation. Furthermore, 
since the Euclidean time $\tau$ becomes an angular type variable with
period 
$\beta$, one is lead also to the identification
\begin{eqnarray}
\gamma_e(y,w)=(y,e^{\frac{i\beta r_+}{\ell^2}}w)\sim (y,w),
\label{eid}
\end{eqnarray}
which is generated by an elliptic element $\gamma_e\in\Gamma$, given by
\begin{eqnarray}
\gamma_e= \left(
\begin{array}{cc}
-e^{\frac{i\beta r_+}{\ell^2}} & 0 \\ 
0 & -e^{-\frac{i\beta r_+}{\ell^2}}
\end{array}
\right).
\label{p1}
\end{eqnarray}
Anyway, requiring the absence of the conical singularity, we get the
relation
\begin{eqnarray}
\frac{\beta r_+}{\ell^2}=2 \pi,
\label{h1456}
\end{eqnarray}
so that $\gamma_e={\textrm{Id}}$, and the period $\beta$, interpreted now
as the inverse of 
the Hawking temperature~\cite{gibb77-15-2752}, is determined to be
\begin{eqnarray}
\beta_H=\frac{2\pi\ell^2}{r_+}.
\label{h1}
\end{eqnarray}
The tree-level Bekenstein-Hawking entropy $S_H$ can also be evaluated, and
is 
given by
\begin{eqnarray}
S_H=\sqrt{\frac{2M}G}\ell=\frac14\frac{2 \pi r_+}G,
\label{bh}
\end{eqnarray}
which is the well known area law for the black hole entropy.

The space-time we are particularly interested in, is the ground state of
the
BTZ black hole, namely the BTZ black hole in the limit of a vanishing
mass;
this space-time is thus described by the line element 
\begin{eqnarray}
ds^2_0=\frac{r^2}{\ell^2}d\tau^2+\frac{\ell^2}{r^2}dr^2+r^2d\varphi^2.
\label{z0}
\end{eqnarray}
This ground state corresponds
also to the zero temperature, zero entropy and zero energy state; moreover 
$r=0$ is a naked singularity in its Lorentzian section \cite{lifs94,henn}
and correspond to a point out of the Euclidean manifold (its geodesical 
distance from the remaining points is infinite).\\ 
By setting 
\begin{eqnarray}
r=\frac{\ell^2}y, \qquad \tau=x_1, \qquad \varphi=\frac{x_2}{\ell},
\end{eqnarray}
we get again the metric of the upper-half model of the hyperbolic space
\begin{eqnarray}
ds^2_0=\frac{\ell^2}{y^2}\left(dy^2+dx_1^2+dx_2^2\right).
\end{eqnarray}
>From (\ref{z0}) and the comment following that equation, 
it is clear that the coordinate 
$\tau$ can be compactified in a circle
with any period $\beta>0$ (in particular $\beta=\infty$) preserving the 
smoothness of the manifold; moreover $\varphi$ has the usual $2\pi$
period. 
In this way, the ground state solution corresponds to the identification
\begin{eqnarray} 
(y,w+\beta +2\pi i\ell) \sim (y,w),
\end{eqnarray}
which is generated by the two parabolic elements
\begin{eqnarray}
\gamma_{p_1}= \left(
\begin{array}{cc}
1 & \beta \\ 
0 & 1
\end{array}
\right),
\qquad \gamma_{p_2}= \left(
\begin{array}{cc}
1 & 2\pi i\ell\\ 
0 & 1
\end{array}
\right)
\:.\label{par}\end{eqnarray}
Thus, our ground state space can be regarded as 
the quotient ${\mathcal{H}}^3_0={\mathbb{H}}^3/\Gamma_0 $, with 
$\Gamma_0$ generated by the two primitive parabolic elements 
$\gamma_{p_1}$ and $\gamma_{p_2}$; one should further notice that, in the 
limit $M \to 0$, the topology of the solution changes and thus the ground 
state case must be considered separately.
 
We finally remind that for negative masses, one gets 
solutions with a naked conical singularity \cite{dese84-152-220} unless 
one arrives at $M=-1$, namely ${\mathbb{H}}^3$, the Euclidean counterpart 
of AdS$^3$; this solution is a permissible one, and can be regarded as a 
bound state \cite{bana92-69-1849}.

\section{The Effective Action for a Scalar Field in the BTZ Ground State.}

In this Section we investigate the spectral properties of a Laplace-like 
operator acting on scalar functions on the non-compact hyperbolic 
manifold ${\mathcal{H}}^3_0$, in order to evaluate the related functional 
determinant, and so the effective action. The BTZ massive case 
has been considered in \cite{mann97-55-3622,byts97u}.
For simplicity, from now on, we put
$\ell=1$ thus $|k|=1/\ell^2=1$ and all the quantities are dimensionless
(the
physical dimensions can be restored by dimensional analysis at the end of
the
calculations).

The heat-kernel related to the Laplace-like operator (see also the
appendix) 
\begin{eqnarray}
L=-\Delta-1,
\end{eqnarray}
is well known, and reads 
\begin{eqnarray}
K_t^{{\mathbb{H}}^3}\!\!\left({\mathbf{x}},{\mathbf{x}}' \vert L\right)=
\frac1{\left(4\pi t\right)^{\frac32}}
\frac{\sigma\!\left({\mathbf{x}},{\mathbf{x}}'\right)}
{\sinh\sigma\!\left({\mathbf{x}},{\mathbf{x}}'\right)}
\exp\left[-\frac{\sigma^2\!\left({\mathbf{x}},{\mathbf{x}}'\right)}{4t}
\right], 
\label{1hhk}
\end{eqnarray}
where the geodesic distance of ${\mathbf{x}}$ from ${\mathbf{x}}'$ in
${\mathbb{H}}^3$ is 
\begin{eqnarray}
\sigma({\mathbf{x}},{\mathbf{x}}')=\cosh^{-1} \left[ 1+\frac{(y-y')^2+
(x_1-x_1')^2+(x_2-x_2')^2}{2yy'}\right],
\label{gd}
\end{eqnarray}
and is usually given in terms of the fundamental invariant of any
pair of points
\begin{eqnarray}
u({\mathbf{x}},{\mathbf{x}}')=\frac12\left[\cosh
\sigma({\mathbf{x}},{\mathbf{x}}')-1\right],  \qquad 
u({\mathbf{x}},{\mathbf{x}})=0.
\label{u}
\end{eqnarray}
Since we are interested in scalar fields propagating in the classical BTZ 
background which are described by the action
\begin{eqnarray}
I\left[\phi\right]=-\frac12\int\!d^3x\,\sqrt
g\left(\nabla_\alpha\phi\nabla^\alpha\phi+m^2\phi^2+
\xi{\mathcal{R}}\phi^2\right),
\label{azio}
\end{eqnarray}
we have to deal with the motion operator
\begin{eqnarray}
L_b=L+b, 
\end{eqnarray}
where $b$ is a constant given in terms of the mass and the gravitational 
coupling of the field,
\begin{eqnarray}
b=1+m^2+\xi\mathcal{R}. \label{B}
\end{eqnarray}
It should be noticed that, in this way, the massless conformally invariant 
case corresponds to the choice $b=1/4$.\\ 
Now, the result (\ref{1hhk}) is trivially generalized for such an
operator, 
and gives
\begin{eqnarray}
K_t^{{\mathbb{H}}^3}\!\!\left({\mathbf{x}},{\mathbf{x}}'\vert L_b\right)=
\frac1{\left(4\pi t\right)^{\frac32}}
\frac{\sigma\!\left({\mathbf{x}},{\mathbf{x}}'\right)}
{\sinh\sigma\!\left({\mathbf{x}},{\mathbf{x}}'\right)}
\exp\left[-tb-\frac{\sigma^2\!\left({\mathbf{x}},{\mathbf{x}}'\right)}{4t}
\right]. 
\label{hhk}
\end{eqnarray}
This Euclidean expression has a Lorentzian counterpart associated with 
Dirichlet boundary condition at spatial infinity, which have to be imposed
because AdS$^3$ is not globally hyperbolic.\\ 
With regard to the heat kernel on $ {\mathcal{H}}^3_0$, we can apply the 
method of images, namely we can write
\begin{eqnarray}
K_t^{{\mathcal{H}}^3_0}\!\left({\mathbf{x}},{\mathbf{x}}'\vert L_b\right) 
& = & \sum_{\gamma_{p}}K^{{\mathbb{H}}^3}_t\!
\left({\mathbf{x}},\gamma_{p}{\mathbf{x}}' \vert L_b\right) \nonumber \\ 
& = & K_t^{{\mathbb{H}}^3}\!\!\left({\mathbf{x}},{\mathbf{x}}'\vert
L_b\right)
+\sum_{\gamma_{p}\neq {\mathrm{Id}}}K^{{\mathbb{H}}^3}_t\!
\left({\mathbf{x}},\gamma_{p}{\mathbf{x}}' \vert L_b\right),
\label{im}
\end{eqnarray}
where the separation between the identity and the non-trivial topological 
contribution has been done, and we have defined
\begin{eqnarray}
\gamma_p=\gamma_{p_1}\cdot\gamma_{p_2}.
\end{eqnarray}
Moreover, notice that the isometry group generated by $\gamma_p$ is
Abelian, so
that the corresponding transformation law for a scalar field reads as
\begin{eqnarray}
\phi(\gamma{\mathbf{x}})=\chi \phi({\mathbf{x}}),
\end{eqnarray}
where $\chi$ is a finite-dimensional unitary representation
(a character) of $\Gamma$.\\ 
So on the diagonal part $({\mathbf{x}}={\mathbf{x}}')$, the heat-kernel 
depends only on $y$, and turns out to be
\begin{eqnarray}
K_t^{{\mathcal{H}}^3_0}({\mathbf{x}},{\mathbf{x}}\vert L_b) & = & 
K_t^{{\mathcal{H}}^3_0}(y\vert L_b) \nonumber \\ 
& = & \frac{e^{-tb}}{(4\pi t)^{\frac32}}+\frac1{(4\pi t)^{\frac32}}
\sum_{{\mathbf{n}} \neq {\mathbf{0}}} 
\frac{\chi_{{\mathbf{n}}}\,\sigma_{{\mathbf{n}}}(y)}{\sinh 
\sigma_{{\mathbf{n}}}(y)}
\exp\left[-tb-\frac{\sigma^2_{{\mathbf{n}}}(y)}{4t}\right],
\label{hkk0}
\end{eqnarray}
with
\begin{eqnarray}
\sigma_{{\mathbf{n}}}(y)=\cosh^{-1}\left[1+\frac{\beta^2 n_1^2+4\pi^2
n_2^2}
{2 y^2}\right]. 
\label{a1}
\end{eqnarray}
It is worth noticing that the Euclidean method selects for the
quantization of 
a scalar field in the BTZ ground state the only boundary condition
(Dirichlet) 
leading to a finite sum over images. Within the Lorentzian methods, since
AdS$^3$ is not globally hyperbolic, also the Neumann and transparent
boundary
conditions can be used (see, for example \cite{lifs94}). However, in
\cite{lifs94} it has been shown that when the mass of the BTZ black hole
goes to zero, only Dirichlet boundary conditions give a regular and smooth
renormalized vacuum expectation value for a scalar field. We will recover
the same
result making use of the $\zeta$-function regularization.

One can now compute the local $\zeta$-function by means of the Mellin 
transform of the heat-kernel (\ref{hkk0}) and then analytically  continue
it 
to the whole complex plane, obtaining
\begin{eqnarray}
\zeta^{{\mathcal{H}}^3_0}\left(s,y\vert L_b\right) & = &
\frac{b^{\frac32-s}}{\left(4\pi\right)^{\frac32}}
\frac{\Gamma\left(s-\frac32\right)}{\Gamma\left(s\right)}+
\frac{b^{\frac34-\frac s2}\,2^{\frac52-s}}{\left(4\pi\right)^{\frac32}
\Gamma\left(s\right)} \nonumber \\ 
& & \times\sum_{{\mathbf{n}}\neq{\mathbf{0}}}\frac{\chi_{{\mathbf{n}}}\,
\sigma^{s-\frac12}_{{\mathbf{n}}}(y)}
{\sinh\sigma_{{\mathbf{n}}}(y)}K_{\frac32-s}(\sqrt b\,
\sigma_{{\mathbf{n}}}(y)).
\label{zetttt}
\end{eqnarray}
The first term in the equation above, is the local $\zeta$-function for
$L_b$
acting on ${\mathbb{H}}^3$, which turns out to be coordinate independent, 
as it should since ${\mathbb{H}}^3$ is a symmetric space. For future 
reference we also report the local $\zeta$-function of the BTZ Euclidean 
section \cite{byts97u}
\begin{eqnarray}
\zeta^{{\mathcal{H}}^3}\left(s,r\vert L_b\right) & = &
\frac{b^{\frac32-s}}{\left(4\pi\right)^{\frac32}}
\frac{\Gamma\left(s-\frac32\right)}{\Gamma\left(s\right)}+
\frac{b^{\frac34-\frac s2}\,2^{\frac52-s}}{\left(4\pi\right)^{\frac32}
\Gamma\left(s\right)} \nonumber \\ 
& & \times\sum_{n\neq 0}\frac{\chi_{n}\,\sigma^{s-\frac12}_{n}(r)}
{\sinh\sigma_{n}(r)}K_{\frac32-s}(\sqrt b\,\sigma_{n}(r)),
\label{btzz}
\end{eqnarray}
where now
\begin{eqnarray}
\sigma_{n}(r)=\cosh^{-1}\left[1+\frac{2r^2}{r_+^2}(\sinh^2 \pi n
r_+)\right].
\label{a1234}
\end{eqnarray}

With regard to the computation of the effective action, one needs the 
analytical continuation of the global $\zeta$-function, obtained by 
performing  the integration over the fundamental domain of the 
diagonal part of the related local quantity.
It is easy to show that the fundamental domain ${\mathcal{F}}_0$ of 
${\mathcal{H}}^3_0$ is non-compact, and that is given as follows 
\begin{eqnarray}
{\mathcal{F}}_0=\left\{ 0\leq y<\infty,\, 0\leq\tau<\beta,\,
0<\varphi<2\pi 
\right\}.
\label{fd}
\end{eqnarray}
This means that the volume $V({\mathcal{F}}_0)=V_0$ of the fundamental
domain, is divergent and we must introduce a regularization; the simplest
one 
consists of limiting the integration in $y$ between~$1/R_0<y<\infty$, 
with $R_0$ large enough.\\ 
Thus we have
\begin{eqnarray}
V_0(R_0)=\int_{1/R_0}^\infty \frac{dy}{y^3}\int_0^{2\pi}
d\varphi \int_0^{\beta}d\tau =\pi \beta R_0^2,
\label{fdv}
\end{eqnarray}
or, in the original coordinates,
\begin{eqnarray}
V_0(R_0)=\int_0^{\beta} d\tau 
\int_0^{2\pi} d\varphi 
\int_{0}^{R_0} r dr=\pi \beta R_0^2.
\label{fdp1}
\end{eqnarray}
In this way, starting from the heat-kernel associated with the
Laplace-like 
operator $L_b$, one has
\begin{eqnarray}
K^{{\mathcal{H}}^3_0}(t\vert L_b)= \frac{V_{0}(R_0)e^{-tb}}{(4\pi
t)^{\frac32}}
+\frac{2\pi\beta e^{-tb}}{(4\pi t)^{\frac32}} \sum_{{\mathbf{n}} \neq 
{\mathbf{0}}}\int_0^\infty\!\frac{dy}{y^{3-\varepsilon}}\,
\frac{\sigma_{{\mathbf{n}}}(y)}{\sinh \sigma_{{\mathbf{n}}}(y)}
\exp\left[{-\frac{\sigma^2_{{\mathbf{n}}}(y)}{4t}}\right],
\label{par61}
\end{eqnarray}
where, as previously remarked, $R$ is the cutoff of the identity volume
element,
and $\varepsilon$ is the parabolic regularization parameter, necessary to 
regularize the divergence associated with the cusp (and  which goes to
zero 
at the end of the calculation). It should be noticed that in~(\ref{par61})
(and from now on), it is assumed that our scalar field obeys to the 
Bose-Einstein statistics ({\emph{i.e.}} $\chi_{{\mathbf{n}}}=1\ \forall 
\,{\mathbf{n}}$).\\ 
Making the change of variable 
\begin{eqnarray}
u=\cosh^{-1} \left[ 1+\frac{\beta^2n_1^2+4\pi^2n_2^2}{y^2}\right],
\end{eqnarray}
one has
\begin{eqnarray}
K^{{\mathcal{H}}^3_0}(t\vert L_b)=\frac{V_{0}(R_0)e^{-tb}}{(4\pi
t)^{\frac32}}+2^{\frac{\varepsilon}2}\pi\beta
E_2\left(1-\frac{\varepsilon}{2}\Big\vert\frac{\beta^2}4,\pi^2\right)
I_{t,b}(\varepsilon),
\label{a2}
\end{eqnarray}
where
\begin{eqnarray}
I_{t,b}(\varepsilon)=\frac{e^{-tb}}{2 (4\pi t)^{\frac32}}
\int_0^\infty\!du\, ue^{-\frac{u^2}{4t}} \left( \cosh u-1 
\right)^{-\frac{\varepsilon}2},
\label{a3}
\end{eqnarray}
and 
\begin{eqnarray}
E_2(s|a_1,a_2)=\sum_{{\mathbf{n}} \neq {\mathbf{0}}}\left( a_1 n_1^2+
a_2 n_2^2 \right)^{-s},
\label{a4}
\end{eqnarray}
is the Epstein $\zeta$-function, which is defined for 
$\mathop{\rm Re}\nolimits s>1$ and can be analytically continued into the 
whole complex plane, its meromorphic continuation having a simple pole 
at $s=1$ and being regular at $s=0$. In particular, one has 
\cite{byts94-9-1569} 
\begin{eqnarray}
E_2(0|a_1,a_2) & = & -1\nonumber, \\ 
E_2'(0|a_1,a_2) & = & \frac12\ln \frac{a_2}{4\pi^2}-2\pi 
\sqrt{\frac{a_1}{a_2}} \zeta(-1)-2 H(2\pi \sqrt{\frac{a_1}{a_2}}),
\label{a5}
\end{eqnarray}
where $H(t)$ is the Hardy-Ramanujan modular function, which is given by
\begin{eqnarray}
H(t)=\sum_{n=1}^\infty \ln \left( 1-e^{-t n} \right)
\:,\label{hr1}\end{eqnarray} 
and satisfies the functional equation
\begin{eqnarray}
H(t)=-\frac{\pi^2}{6t}-\frac{1}{2} \ln \left( \frac{t}{2 
\pi}\right)+\frac{t}{24}+H(\frac{4\pi^2}{t}).
\label{hr12}
\end{eqnarray}
Making use of the 
Epstein functional equation with $a_1=(\beta/2)^2$, $a_2=\pi^2$, one has
\begin{eqnarray}
E_2\left(1-\frac{\varepsilon}2\Big\vert\frac{\beta^2}4,\pi^2\right)=
\frac2{\beta\pi^{\varepsilon}}\frac{2\Gamma\left(\frac{\varepsilon}2\right)}
{\Gamma\left(1-\frac{\varepsilon}2\right)}
E_2\left(\frac{\varepsilon}2\Big\vert\frac4{\beta^2},\frac1{\pi^2}\right)
+O(\varepsilon),
\label{a6}
\end{eqnarray}
so that, after a first order expansion,
\begin{eqnarray}
K^{{\mathcal{H}}^3_0}(t\vert L_b) & = & \frac{V_{0}(R_0)e^{-tb}}{(4\pi
t)^{\frac32}}-\frac{4\pi}{\varepsilon}I_{t,b}(0)-4\pi\left[I'_{t,b}(0)
+I_{t,b}(0)G(\beta)\right]+O(\varepsilon),
\label{a61}
\end{eqnarray}
where
\begin{eqnarray}
I_t'(0)=-\frac{e^{-t}}{16\sqrt\pi t}\int_0^\infty du\, u 
e^{-\frac{u^2}{4t}}  \ln \left( \cosh u-1 \right),
\label{a7}
\end{eqnarray}
and, finally,
\begin{eqnarray}
G(\beta)=\frac32\ln2+\ln\pi-C+\frac{4\pi^2}{\beta}\zeta(-1)+
2H(\frac{4\pi^2}{\beta}),
\end{eqnarray}
($C$ is the Euler-Mascheroni constant).\\ 
So, besides the divergence of the volume 
(non-compact manifold) controlled by  $R$, one has another divergence due
to 
the continuum spectrum associated with the cusp, namely the pole at
$\varepsilon=0$. It turns out that this singularity appears also in the 
spectral representation of the heat-kernel trace and it may be removed by 
means of suitable  definition of the trace, as in the case of 
non-compact hyperbolic manifold with finite volume 
(see, for example \cite{byts97-30-3543} and references quoted therein). 
Thus, one has
\begin{eqnarray}
K^{{\mathcal{H}}^3_0}(t\vert L_b) & = & \frac{V_{0}(R_0)e^{-tb}}{(4\pi
t)^{\frac32}}-4\pi\left[I'_{t,b}(0)+I_{t,b}(0)G(\beta)\right]+O(\varepsilon),
\label{a62}
\end{eqnarray}

As a consequence of the obtained results, 
one can now compute the global $\zeta$-function associated
with our operator $L_b$, finding 
\begin{eqnarray}
\zeta\left(s\vert L_b\right) & = &
\frac{V_{0}(R_0)b^{\frac32-s}}
{\left(4\pi\right)^{\frac32}}
\frac{\Gamma\left(s-\frac32\right)}{\Gamma\left(s\right)}-
\frac{G\left(\beta\right)b^{\frac12-s}}{\sqrt{4\pi}}
\frac{\Gamma\left(s-\frac12\right)}{\Gamma\left(s\right)}\nonumber \\ 
& & +\frac{b^{\frac12-s}}{\sqrt{4\pi}}
\frac{\Gamma\left(s-\frac12\right)}{\Gamma\left(s\right)}
\left[-\frac12\log b+\frac{\Psi\left(s-\frac12\right)-C}2\right]\nonumber
\\ 
& & +\frac{2^{-s}\,b^{\frac34-s}}{\sqrt{2\pi}\Gamma\left(s\right)}
\int_0^\infty\!dz\,z^{s-\frac12}K_{\frac32-s}(\sqrt bz)
\left[\log\left(\cosh z-1\right)-2\log\frac z2\right],
\label{bbbbbb12}
\end{eqnarray}
where the last integral is convergent.\\ 
It should be noticed that, due to the presence of parabolic elements, the 
meromorphic structure of this $\zeta$-function contains double poles at $ 
s=1/2-k,\,\,k=0,1,2,\ldots$; moreover this~$\zeta$-function is analytic in 
$s=0$, and its derivative reads
\begin{eqnarray}
\ln \left(\det L_b\right) & = & -\zeta'(0\vert L_b) \nonumber \\ 
& = & -\frac{ 
V_{0}(R_0)b^{\frac32}}{6\pi}-G(\beta)\sqrt b-F_b,
\label{io}
\end{eqnarray}
where $F_b$ is a constant (independent from $\beta$) given by
\begin{eqnarray}
F_b & = & 
\sqrt b\left[\frac12\log b+C+\log2-1\right] \nonumber \\ 
& & +\frac{b^{\frac34}}{\sqrt{2\pi}}
\int_0^\infty\!dz\,z^{-\frac12}K_{\frac32}(\sqrt bz) 
\left[\log\left(\cosh z-1\right)-2\log\frac z2\right].
\end{eqnarray}

\section{The First Quantum Correction to the Entropy of the BTZ Black
Hole.}

The first on-shell quantum correction to the Bekenstein-Hawking entropy
may be 
computed within the Euclidean semiclassical approximation 
\cite{gibb77-15-2752} and we shall follow this approach in this Section. 
A pure gravitational quantum correction to the BTZ entropy has been
presented 
in \cite{byts97u}, making use of Chern-Simons representation of the  
3-dimensional gravity \cite{witt88-311-46}. Very recently in
\cite{brotz98} the
first quantum correction to the entropy and the back reaction of the BTZ
black
hole also have been studied.
Here, for the sake of simplicity, we assume that the quantum degrees of 
freedom of the massive black hole are represented by the quantum 
scalar field (described as usual by the 
action (\ref{azio})) propagating outside the black hole \cite{thoof}, 
and we shall make use of the results of \cite{byts97u} as well as the ones 
obtained in  Sect.~III.\\ 
Recall that within the Euclidean approach, the one-loop approximation
gives,
for the partition function in the BTZ background,
\begin{eqnarray}
Z_{BTZ}=e^{-I_M} \left(\det L_b\right)_M^{-1/2},
\label{part0}
\end{eqnarray}
where $I_M$ is the classical action related to the massive BTZ solution 
(see, for example, \cite{byts97u}). It reads
\begin{eqnarray}
I_M=I_{BTZ}+B_{BTZ},
\label{bbbbbb}
\end{eqnarray}
in which $I_{BTZ}$ is the Hilbert-Einstein action, while $B_{BTZ}$ is the
usual
boundary term which depends on the extrinsic curvature at large spatial
distance. We remind that the total classical action is divergent;
the geometry is non-compact and one  has to introduce the reference 
background ${\mathcal{H}}_0^3$ at least at the tree level 
\cite{hawk96-13-1487}, and the related  volume cutoffs $R$ and $R_0$. 
Thus, one may also consider the related ground state partition function 
\begin{eqnarray}
Z_{BTZ_0}=e^{-I_0} \left(\det L_b \right)_0^{-1/2},
\label{part00}
\end{eqnarray}
where $I_0$ is the classical action related to the massless BTZ solution, 
given by 
\begin{eqnarray}
I_{0}=I_{BTZ_0}+B_{BTZ_0}.
\label{bbbbbb1}
\end{eqnarray}
A simple but crucial observation is that, in order to recover the tree 
level Bekenstein-Hawking entropy, one may introduce the ``relative''
partition function
\begin{eqnarray}
Z_r=\frac{Z_{BTZ}}{Z_{BTZ_0}}=\left[ \frac{\left(\det
L_b\right)_0}{\left(\det
L_b\right)_M} \right]^{\frac12} 
e^{-(I_M-I_{0})}.
\label{rp}
\end{eqnarray} 
With this proposal, the two boundary terms of the classical 
contribution cancel for large $r$ and the difference of
the on-shell Euclidean classical actions leads to \cite{byts97u}, 
\begin{eqnarray}
I_M-I_{0}=I_{BTZ}-I_{BTZ_0}=-\frac{2}{\pi} \left( V(R)-V_0(R_0) \right) 
\to -2 \pi r_+.
\label{bh1}
\end{eqnarray}
Restoring the correct physical dimension, it is easy to show that 
the on-shell tree-level partition function $ Z^{(0)}$ becomes
\begin{eqnarray}
\ln Z^{(0)}=\frac{\pi^2 r_+}{4 \pi G},
\label{zoo}
\end{eqnarray}
and this leads to the semiclassical Bekenstein-Hawking entropy
\begin{eqnarray}
S^{(0)}=S_H=\left( r_+\frac{\partial}{\partial r_+}+1 \right) \ln Z^{(0)}=
\frac{1}{4}\frac{2 \pi r_+}{G}.
\label{bh11}
\end{eqnarray}
Furthermore, concerning the regularization of the ratio of the two 
functional determinants (representing the quantum corrections), 
our proposal implements the correct mathematical procedure,
that is necessary when one is dealing with functional determinants of 
elliptic operators on non-compact manifold (see \cite{mull98}).
In fact, in our case the  manifolds are non-compact and a 
volume regularization (as the one previously introduced) must be used. 
Thus, we have
\begin{eqnarray}
\ln Z_r=2\pi r_+ +
\frac{1}{2}\ln\left(\det L_b\right)_0-\frac{1}{2}\ln\left(\det
L_b\right)_M.
\label{ziii}
\end{eqnarray} 

In the case of scalar fields, one can compute the functional determinants 
in the BTZ background. Using the $\zeta$-function regularization and the 
volume cutoff $R$ and $R_0$, as well as (\ref{io}) with $\beta=\beta_H$, 
one gets
\begin{eqnarray}
\ln Z_r(R)=\frac{\pi r_+}{4 G}+\frac{b^{\frac32}V(R)}{12
\pi}-\frac{1}{2}\ln 
{{\mathcal{Z}}_0}(2)-\frac{b^{\frac32}V_0(R_0)}{12 \pi}-\frac{F_b}2
-\frac{\sqrt bG(r_+)}2,
\label{bh123}
\end{eqnarray}
where we have introduced the function
\begin{eqnarray}
\ln {\mathcal{Z}}_0(2)=\sum_{n=1}^\infty\frac1n\left(e^{\pi nr_+
\frac{\sqrt b+1}2}-e^{\pi nr_+\frac{\sqrt b-1}2}\right)^{-2}.
\end{eqnarray}
Now we can remove the volume cutoff, taking the limit $R \to \infty$. In 
this way the horizon divergences cancel out and the finite result can be 
written as
\begin{eqnarray}
\ln Z_r=\frac{\pi r_+}{4 G}+h(r_+),
\label{bh1234}
\end{eqnarray}
where
\begin{eqnarray}
h(r_+)=-\frac{1}{2}\ln {\cal Z}_0(2)-\frac{b^{\frac32}\pi r_+}{12}
-\frac{F_b}2-\frac{\sqrt bG(r_+)}2.
\label{hhhp}
\end{eqnarray}
Here $G$ can be identified with an effective Newton constant and we 
stress that within this approach, the horizon divergences have been 
dealt with without an ultraviolet renormalization of it. 
This finite relative one-loop effective action may be thought to describe
an 
effective classical geometry belonging to the same class of the non
rotating 
BTZ black hole solution. This stems from the results contained in 
\cite{henn}, where it has been shown that the constraints for pure gravity 
have an unique solution. As a consequence, one may introduce a new
effective 
radius by means of 
\begin{eqnarray}
\ln Z_r=\frac{\pi R_+}{4 G},
\label{bbb2}
\end{eqnarray}
where
\begin{eqnarray}
R_+=r_+ + \frac{4G}{\pi} h(r_+),
\label{r}
\end{eqnarray}
mimicking in this way the back reaction of the quantum gravitational 
fluctuations. As a consequence, the new  entropy  is given by an effective 
Bekenstein-Hawking term, namely 
\begin{eqnarray}
S^{(1)}=\frac{1}{4}\frac{ 2\pi R_+}{G}.
\label{effent}
\end{eqnarray}
One can evaluate the asymptotic behavior of the quantity $h(r_+) $ for 
$r_+ \to \infty$ and $r_+ \to 0$, and then 
obtain the effective radius. Notice that $H(r_+)$ and
$\ln{\mathcal{Z}}_0(2)$ 
are exponentially small for large  $r_+$. Thus, being $c$ a numerical
factor, 
we find
\begin{eqnarray}
R_+ \simeq r_+ +c \sigma Gr_+,
\label{bnm}
\end{eqnarray}
where $c \sigma G r_+$ are quantum corrections, which may be small since 
$G$ is the inverse  of the Planck length. On the other hand, for small
$r_+$
one has
\begin{eqnarray}
R_+ \simeq  r_++\frac{4G}{\pi} \left[ \frac{\sigma^2}{16r_+^2} +
c_1\frac{\sigma}{ r_+}+O( \ln \left( \frac{r_+}{\sigma 
\pi}\right)) \right],
\label{bnm1}
\end{eqnarray}
where $c_1$ is another numerical factor.\\ 
One can see that for $r_+$ sufficiently small the effective radius 
becomes larger and positive. This means that $R_+$ (as a function of
$r_+$) 
reaches a minimum for a suitable $r_+$. This result is in qualitative
agreement 
with a very recent computation of the off-shell quantum 
correction to the entropy due to a scalar field in the BTZ background 
\cite{mann97-55-3622} and for the pure gravitational case \cite{byts97u}.
In particular, it appears that the quantum gravitational 
corrections could become more and more important as soon as the
evaporation 
process continues and thus they cannot be neglected. This qualitative 
picture does not take into account the back reaction. In order to do this, 
one must compute the vacuum expectation value of the stress tensor.

\section{The Vacuum Expectation Value of the Stress Tensor.}  
 
In this Section, we shall compute the expectation value of the square of 
a quantum scalar field and its associated stress tensor expectation 
value on the black hole background. The latter will be used in the
computation 
of the back reaction, by solving the semiclassical Einstein equations
\begin{eqnarray}
{\mathcal{R}}_{\mu \nu}-\frac12g_{\mu \nu}{\mathcal{R}}+\Lambda g_{\mu
\nu}=
8\pi G \langle T_{\mu \nu}\rangle.
\label{semi}
\end{eqnarray}  
With regard to this issue, it is worthwhile noticing that customary used 
methods based on the correct behavior of the Green function to 
pick out a particular temperature for the thermal state are useless
in the present contest. Indeed, such methods consider the behavior
of the Green function when an argument belongs to some particular
relevant point of the manifold, in particular points of the
event horizons \cite{hahem}, and require a  correct scaling limit
for short distances as well as Hadamard's behavior. In the present case,
no
horizon appears and the singular points at $r=0$ are not in the manifold
as far as the Euclidean section of it is concerned. This is
because any geodesic falling into these points spents an infinite amount
of affine parameter. In the Lorentzian section, some points at $r=0$,
which are singular \cite{lifs94,henn}, belong to the manifold because, 
for instance, some time-like geodesics can reach
such points in a finite period of proper time. Anyhow, in this case, the 
set of points at $r=0$
represents a naked singularity and the use of the principles above
for arguments of the Green functions fixed at $r=0$ seems to be very 
problematic. On the other hand, in the Euclidean section, the request of 
absence of the conical singularities, does not select any temperature.
For these reasons we shall deal with all  possible values 
of the inverse temperature $\beta>0$, so that
one has to consider the full (parabolic) isometry group of the ground
state 
(whereas, in the case of the zero temperature state, one should deal only 
with the element $\gamma_{p_2}$ of (\ref{par})).

Let us now consider a  non-minimally coupled  scalar field $\phi$,
described 
by the action~(\ref{azio}). 
We recall that within the $\zeta$-function regularization,
one has \cite{more,moreL}
\begin{eqnarray}
\langle\phi^2({\mathbf{x}})\rangle=\lim_{s \to 0}
\left[ \zeta(s+1,{\mathbf{x}}\vert L_b)+
s\zeta'(s+1,{\mathbf{x}}\vert L_b) \ln\mu^2 \right].
\label{m1}
\end{eqnarray}
The substantial equivalence between the formula above and the 
result of point splitting procedure has been analyzed in \cite{moreL}.
In $D=3$, the local $\zeta$-function is regular at $s=1$ and the
dependence 
on the scale parameter $\mu^2$ drops out; thus one has
\begin{eqnarray}
\langle\phi^2(y)\rangle & = & \frac{\sqrt
b}{(4\pi)^{\frac32}}\Gamma\left(-\frac12\right)+\frac{b^{\frac14}2^{\frac32}}
{(4\pi)^{\frac32}}\sum_{{\mathbf{n}}\neq{\mathbf{0}}}
\frac{\sqrt{\sigma_{{\mathbf{n}}}(y)}}
{\sinh\sigma_{{\mathbf{n}}}(y)}K_{\frac12}(\sqrt
b\sigma_{{\mathbf{n}}}(y)) 
\nonumber \\ 
& = &\frac{-\sqrt b}{4\pi}+\frac1{4\pi}\sum_{{\mathbf{n}}\neq{\mathbf{0}}}
\frac{e^{-\sqrt
b\sigma_{{\mathbf{n}}}(y)}}{\sinh\sigma_{{\mathbf{n}}}(y)}. 
\label{m2}
\end{eqnarray}
Notice that the ${\mathbb{H}}^3$ case corresponds  to the 
first term in the above equation and it turns out that the contribution is 
negative, namely one has
\begin{eqnarray}
\langle\phi^2(y)\rangle^{{\mathbb{H}}^3}=-\frac{\sqrt b}{4\pi}.
\label{nnn}
\end{eqnarray}
The second term can be referred to as the ``topological term'' and may be 
rewritten noticing that
\begin{eqnarray}
\sigma_{{\mathbf{n}}}(y)=
\ln\left(1+C_{{\mathbf{n}}}+\sqrt{C_{{\mathbf{n}}}^2+
2C_{{\mathbf{n}}}}\right),
\label{a11}
\end{eqnarray}
where we have introduced the function
\begin{eqnarray}
C_{{\mathbf{n}}}({\mathbf{x}},{\mathbf{x}}')=\frac{(y-y')^2+(x_1-x'_1-
\beta n_1)^2+(x_2-x'_2-2\pi n_2)^2}{2y y'},
\label{coff}
\end{eqnarray}
that on the diagonal reads
\begin{eqnarray}
C_{{\mathbf{n}}}(y)=\frac{2b^2_{{\mathbf{n}}}}{y^2}, 
\qquad b^2_{{\mathbf{n}}}=\frac{\beta^2n_1^2}4+\pi^2n_2^2.
\label{a22}
\end{eqnarray}
A direct computation of the field fluctuation as a function of 
$C_{{\mathbf{n}}}$, leads to
\begin{eqnarray}
\langle\phi^2(y)\rangle=-\frac{\sqrt b}{4\pi}
+\sum_{{\mathbf{n}}\neq{\mathbf{0}}} {\mathcal{H}}(C_{{\mathbf{n}}}(y)),
\label{nnno}
\end{eqnarray}
with
\begin{eqnarray}
{\mathcal{H}}(C_{{\mathbf{n}}})=\frac{2^{\sqrt b-3}}{\pi}
\left( \frac{1}{\sqrt C_{{\mathbf{n}}}}-
\frac{1}{\sqrt{C_{{\mathbf{n}}}+2}}\right)\left( \sqrt{C_{{\mathbf{n}}}}+
\sqrt{C_{{\mathbf{n}}}+2} \right)^{1-2\sqrt b}.
\label{h}
\end{eqnarray}
The series which appears in the right hand side of (\ref{nnno}) is 
convergent as soon as $b>0$.\\ 
A similar computation in the BTZ case, namely $M>0$, yields the same 
result, but with
\begin{eqnarray}
C_n({\mathbf{x}},{\mathbf{x}}')=
\frac{(N^{\frac{n}2}y-N^{-\frac{n}2}y')^2+
(N^{\frac{n}2}x_1-N^{-\frac{n}2}x'_1)^2+
(N^{\frac{n}2}x_2-N^{-\frac{n}2}x'_2)^2}{2y y'},
\label{coffbtz}
\end{eqnarray}
in place of $C_{\mathbf{n}}$,
where $\ln N=2\pi r_+$, and on the diagonal 
\begin{eqnarray}
C_n(r)=\frac{r^2}{r^2_+}\sinh^2 2\pi n r_+.
\label{btzc}
\end{eqnarray}
In particular, in the massless conformally invariant case, one has 
\begin{eqnarray}
\langle\phi^2(r)\rangle^{\mathrm{BTZ}}=-\frac{1}{8\pi}
+\frac{1}{2 \sqrt 2 \pi}\sum_{n=1}^\infty \left( \frac{1}{\sqrt 
C_n}-\frac{1}{\sqrt{C_n+2}} \right),
\label{cibtz}
\end{eqnarray}
in agreement with the result reported in \cite{lifs94}.

As far as  the expectation value of the stress tensor related to 
the field $\phi$ is concerned, in $D=3$ we have \cite{more1,ultimo}
\begin{eqnarray}
\langle T_{\mu \nu}({\mathbf{x}})\rangle=\zeta_{\mu \nu}
(1,{\mathbf{x}}\vert L_b),
\label{m3}
\end{eqnarray}
where the right hand side of the  equation above is defined (in the sense
of 
the analytical continuation) as
\begin{eqnarray}
\zeta_{\mu \nu}(s,{\mathbf{x}}\vert L_b)=\sum_n \lambda_n^{-s} T_{\mu \nu}
(\phi^*_n,\phi_n),
\label{m4}
\end{eqnarray}
with $\phi_n$ representing the eigenfunctions of the Laplace-like
operator 
$L_b$ and $T_{\mu \nu}(\phi_n^*,\phi_n)$ being the  classical stress
tensor 
evaluated on the modes. The latter is defined as
\begin{eqnarray}
T_{\mu \nu}(\phi^*,\phi)
=\frac{2}{\sqrt g}\frac{\delta I[\phi^*,\phi]}{\delta g^{\mu \nu}},
\label{st}
\end{eqnarray}
where $I[\phi^*,\phi]_{\phi^*=\phi}$ is the associated classical action.\\ 
Furthermore, it is possible to show that~\cite{more1,ultimo}
\begin{eqnarray}
\zeta_{\mu \nu}(s,{\mathbf{x}}\vert{L_b})=L_{\mu \nu}
\zeta(s,{\mathbf{x}}\vert L_b)-\frac12g_{\mu \nu}
\zeta(s-1,{\mathbf{x}}\vert L_b)+
\bar{\zeta}_{\mu \nu}(s,{\mathbf{x}}\vert L_b),
\label{m5}
\end{eqnarray}
where
\begin{eqnarray}
L_{\mu \nu}= \xi{\mathcal{R}}_{\mu \nu}+\left( \xi-\frac14\right)
g_{\mu \nu}\Delta-\xi \nabla_\mu \nabla_\nu,
\label{m7}
\end{eqnarray}
and, again in the sense of the analytical continuation,
\begin{eqnarray}
\bar{\zeta}_{\mu \nu}(s,{\mathbf{x}}\vert L_b)=
\frac12\sum_n \lambda_n^{-s}\left(\partial_{\mu}\phi^*_n 
\partial_\nu \phi_n +\partial_{\nu}\phi^*_n \partial_\mu \phi_n \right).
\label{m67}
\end{eqnarray}
As a result, in $D=3$, since $\zeta(0,{\mathbf{x}}\vert L_b)=0$, one has
\begin{eqnarray}
\langle T_{\mu \nu}({\mathbf{x}})\rangle=\lim_{s \to 1}\left[L_{\mu \nu}
\zeta(s,{\mathbf{x}}\vert L_b)+\bar{\zeta}_{\mu 
\nu}(s,{\mathbf{x}}\vert L_b)\right].
\label{m6}
\end{eqnarray}
Now, recalling that we are dealing with  quotient manifolds 
${\mathbb{H}}^3\!/\Gamma$, the image sum method can be applied. 
In general, our $\zeta$-functions are so the sum of two contributions,
namely
\begin{eqnarray}
\zeta(s,{\mathbf{x}},{\mathbf{x}}'\vert 
L_b)=\zeta^{{\mathbb{H}}^3}(s,{\mathbf{x}},{\mathbf{x}}'\vert
L_b)+\zeta^{\Gamma}(s,{\mathbf{x}},{\mathbf{x}}'\vert L_b).
\label{ima}
\end{eqnarray}
Thus, in our case, the expectation value of the stress tensor splits 
in the sum of the related contributions
\begin{eqnarray}
\langle T_{\mu \nu}({\mathbf{x}})\rangle=
\langle T_{\mu \nu}({\mathbf{x}})\rangle^{{\mathbb{H}}^3}
+\langle T_{\mu \nu}({\mathbf{x}})\rangle^{\Gamma}.
\label{ttt}
\end{eqnarray}
Let us compute the first contribution. Now 
$\zeta^{{\mathbb{H}}^3}(s,{\mathbf{x}}\vert L_b)$ is independent from 
${\mathbf{x}}$, and thus
\begin{eqnarray}
\lim_{s \to 1} L_{\mu \nu}\zeta^{{\mathbb{H}}^3}(s,{\mathbf{x}}\vert L_b)=
-2 \xi \zeta^{{\mathbb{H}}^3}(1,{\mathbf{x}}\vert L_b) 
g_{\mu \nu}^{{\mathbb{H}}^3}
=\frac{\xi \sqrt b}{2 \pi}g_{\mu \nu}^{{\mathbb{H}}^3},
\label{hhh}
\end{eqnarray}
$g_{\mu \nu}^{{\mathbb{H}}^3}$ being the ${\mathbb{H}}^3$ metric.
Furthermore, making use of the eigenfunctions reported in the appendix,
one easily finds the following analytical continuation
\begin{eqnarray}
\bar{\zeta}_{\mu \nu}^{{\mathbb{H}}^3}(s,{\mathbf{x}}\vert L_b)=
\frac{1}{12\pi^2 \Gamma(s)}
\left[\Gamma\left(\frac32\right)\Gamma\left(s-\frac32\right)b^{\frac32-s}+
\Gamma\left(\frac52\right)\Gamma\left(s-\frac52\right)b^{\frac52-s}
\right]
g_{\mu \nu}^{{\mathbb{H}}^3}.
\label{m72}
\end{eqnarray}
In this way we got the result
\begin{eqnarray}
\langle T_{\mu \nu}({\mathbf{x}})\rangle^{{\mathbb{H}}^3}= 
\frac{\sqrt b}{4 \pi}
\left(\frac{b-1}{3}+2\xi \right) g_{\mu \nu}^{{\mathbb{H}}^3}
=-\frac{m^2}3\langle\phi^2(y)\rangle^{{\mathbb{H}}^3}
g_{\mu \nu}^{{\mathbb{H}}^3},
\label{m8}
\end{eqnarray}
with the related trace
\begin{eqnarray}
g^{\mu \nu}_{{\mathbb{H}}^3}\langle T_{\mu \nu}({\mathbf{x}})
\rangle^{{\mathbb{H}}^3}=-m^2 \langle\phi^2(y)\rangle^{{\mathbb{H}}^3}
\label{llll}
\end{eqnarray}
in agreement with the general formula \cite{more}\footnote{The coefficient
$1/2\xi_D$ which appears in (13) of \cite{more} is missprinted, and
has to be corrected into $1/(4\xi_D-1)$.}
\begin{eqnarray}
g^{\mu \nu}\langle T_{\mu \nu}({\mathbf{x}})\rangle=
\zeta(0,{\mathbf{x}}\vert A)-
\left(m^2+\frac{\xi-\xi_D}{4\xi_D -1}\Delta \right) 
\langle\phi^2({\mathbf{x}})\rangle.
\label{moe}
\end{eqnarray}
In particular, in the massless conformally coupled case  
one has $\langle T_{\mu \nu}({\mathbf{x}})
\rangle^{{\mathbb{H}}^3}=0$, in agreement with the fact that
${\mathbb{H}}^3$ 
is a homogeneous symmetric space, and that the conformal anomaly vanishes
in 
odd dimensions.  

For the topological non trivial part  
$\langle T_{\mu \nu}({\mathbf{x}})\rangle^{\Gamma}$, it is convenient to 
proceed as follows.\\ 
Making use of the  of the eigenvalues equation for the scalar
eigenfunctions 
\begin{eqnarray}
L_b\phi_n = \lambda_n \phi_n,
\end{eqnarray}
and the background metric form, a standard calculation for the stress
tensor evaluated on the modes (\ref{st}) leads to
\begin{eqnarray}
2T_{\mu \nu}(\phi_n^*,\phi_n)({\mathbf{x}}) & = &
(1-2\xi)\left( \nabla_\mu \phi^*_n 
\nabla_\nu  \phi_n+ \phi^*_n \nabla_\mu \nabla_\nu  \phi_n \right) 
\nonumber \\ 
& & +(2\xi-\frac{1}{2})g_{\mu \nu}\left[ \left( \nabla |\phi_n|
 \right)^2 +\phi^*_n \Delta \phi_n \right]  -\frac{m^2 }{3} g_{\mu \nu} 
|\phi_n|^2 \nonumber \\ 
& & +\left( \frac{1}{3}g_{\mu \nu} \phi^*_n
 \Delta \phi_n-\phi_n^* \nabla_\mu \nabla_\nu  \phi_n 
\right) \nonumber\\ 
& & +\{\phi_n \rightarrow \phi^*_n, \phi^*_n \rightarrow \phi_n \} -
\frac{\lambda_n g_{\mu\nu}}{3} |\phi_n|^2.
\label{mmmm1}
\end{eqnarray}
Then, we can make use of (\ref{m3}), noticing that the last term
in the equation above cannot product a contribution to the final stress 
tensor because it should be proportional to
\begin{eqnarray}
g_{\mu\nu}\left(\zeta(0,{\mathbf{x}}\vert L_b) -
\zeta^{{\mathbb{H}}^3}(0,{\mathbf{x}}\vert L_b)\right) = 0,
\end{eqnarray}
that vanishes since $D=3$ is odd so that both the $\zeta$ functions above 
vanishes for $s=0$ (remember that there is no conformal anomaly in 
odd-dimensional space times). Moreover, following the analysis contained
in
\cite{ultimo}, it is possible to prove that the function
$\zeta^{\Gamma}_{\mu\nu}(1,{\mathbf{x}}\vert L_b)$
of the topological non-trivial part of the stress tensor 
can be computed as the coincidence limit of the corresponding off-diagonal
$\zeta$-function. This is because the corresponding series
does not contain the identity element which gives rise to divergences. 
In general, the equivalence drops out for this element just
because of the existence of a singularity at the coincidence limit.
In practice, concerning the non-trivial topological part of the stress
tensor,
from (\ref{mmmm1}) and (\ref{m3}), one finds that it reduces to
\begin{eqnarray}
\langle T_{\mu \nu}({\mathbf{x}})\rangle^{\Gamma}
=(1-2\xi)A_{\mu \nu}+\left(2\xi-\frac12\right) g_{\mu \nu} 
A +\frac{1}{3}g_{\mu \nu} B-B_{\mu \nu}-\frac{m^2}{3}g_{\mu \nu} 
\zeta^{\Gamma}(1,{\mathbf{x}}\vert L_b),
\label{trca}
\end{eqnarray}
where we have defined 
\begin{eqnarray}
A_{\mu \nu}=\lim_{{\mathbf{x}}' \to {\mathbf{x}}}\frac12
\left[(\nabla_\mu \nabla'_\nu+\nabla'_\mu \nabla_\nu)+ 
(\nabla_\mu \nabla_\nu+\nabla'_\mu \nabla'_\nu) \right] 
\zeta^{\Gamma}(1,{\mathbf{x}},{\mathbf{x}}'\vert L_b),
\label{tmn1}
\end{eqnarray}
with $A=g^{\mu \nu} A_{\mu \nu}$, and
\begin{eqnarray}
B_{\mu \nu}=\lim_{{\mathbf{x}}' \to {\mathbf{x}}}\frac12
\left[(\nabla_\mu \nabla_\nu+\nabla'_\mu \nabla'_\nu) \right] 
\zeta^{\Gamma}(1,{\mathbf{x}},{\mathbf{x}}'\vert L_b),
\label{tmn}
\end{eqnarray}
with $B=g^{\mu \nu} B_{\mu \nu}$.
Moreover, since
\begin{eqnarray}
\zeta^{\Gamma}(1,{\mathbf{x}},{\mathbf{x}}'\vert L_b)=
\sum_{{\mathbf{n}}\neq{\mathbf{0}}}
{\mathcal{H}}(C_{{\mathbf{n}}}({\mathbf{x}},{\mathbf{x}}')),
\label{zoffo}
\end{eqnarray}
a direct calculation in the coordinate system $(y,x_1,x_2)$ leads to
\begin{eqnarray}
A_{\mu\nu} & = & \sum_{{\mathbf{n}}\neq{\mathbf{0}}}
\left[\left(\frac{8b_{{\mathbf{n}}}^4
{\mathcal{H}}''}{y^6}+\frac{2b_{{\mathbf{n}}}^2
{\mathcal{H}}'}{y^4}\right)\delta_{\mu0}\delta_{\nu0}+
\frac{2b_{{\mathbf{n}}}^2{\mathcal{H}}'}{y^2}
g^{{\mathbb{H}}^3}_{\mu\nu}\right], \\*
A & = & \sum_{{\mathbf{n}}\neq{\mathbf{0}}}
\left[\frac{8b_{{\mathbf{n}}}^4{\mathcal{H}}''}
{y^4}+\frac{8b_{{\mathbf{n}}}^2{\mathcal{H}}'}
{y^2}\right], \\ 
B_{\mu\nu} & = & \sum_{{\mathbf{n}}\neq{\mathbf{0}}}
\left[\left(\frac{4b_{{\mathbf{n}}}^4}{y^6}
\delta_{\mu0}\delta_{\nu0}
+\frac{\beta^2n_1^2}{y^4}\delta_{\mu 1}\delta_{\nu 1}
+\frac{4\pi^2n_2^2}{y^4}\delta_{\mu2}\delta_{\nu2}\right)
{\mathcal{H}}''\right. \nonumber \\ 
& & +\left.\left(g^{{\mathbb{H}}^3}_{\mu\nu}+\frac{2b_{{\mathbf{n}}}^2}
{y^2}g^{{\mathbb{H}}^3}_{\mu\nu}\right){\mathcal{H}}'\right], \\ 
B & = & \sum_{{\mathbf{n}}\neq{\mathbf{0}}}
\left[\left(\frac{4b_{{\mathbf{n}}}^4}{y^4}
+\frac{4b_{{\mathbf{n}}}^2}{y^2}\right){\mathcal{H}}''+
\left(3+\frac{6b_{{\mathbf{n}}}^2}{y^2}\right){\mathcal{H}}'\right], 
\end{eqnarray}
where the prime means derivatives with respect to $C_{{\mathbf{n}}}$.\\
Summarizing, we have found that, the complete renormalized stress
tensor is that written in the right hand side of (\ref{ttt}) where the
former  term is given in (\ref{m8}) taking account of (\ref{nnn}), and the
latter is given in (\ref{trca}) taking account of (\ref{h}), (\ref{zoffo})
and the expressions for ${A_{\mu\nu},A,B_{\mu\nu},B}$ written above.
Moreover notice that the dependence on $\xi$ and $m^2$ arises
only from ${\mathcal{H}}$ and its derivatives, and is given by (\ref{h}); 
the $\beta$ dependence is instead due to $b_{{\mathbf{n}}}$ and 
${\mathcal{H}}$ and is given by (\ref{a22}) and (\ref{h}).\\ 
In the zero temperature case one has the same result, but replacing 
$b^2_{{\mathbf{n}}}$ with $\pi^2n_2^2$, dropping the term proportional to 
$\beta$ in $B_{\mu \nu}$, and considering only the sum over $n_2$.\\ 
With regard to the stress tensor trace one finally has
\begin{eqnarray}
g^{\mu \nu}\langle T_{\mu \nu}({\mathbf{x}})\rangle^{\Gamma}=
4 \left( \xi-\frac18\right) 
A-m^2\zeta^{\Gamma}(1,{\mathbf{x}},\vert L_b),
\label{stt}
\end{eqnarray} 
so that the total contribution reads
\begin{eqnarray}
g^{\mu \nu}\langle T_{\mu \nu}({\mathbf{x}})\rangle & = & \langle T\rangle
=
4 \left( \xi-\frac{1}{8}\right) A
-m^2 \langle \phi^2({\mathbf{x}})\rangle \nonumber\\ 
& = & \left[2\left(\xi -\frac{1}{8} \right)\Delta -m^2
\right]\langle\phi^2
({\mathbf{x}})\rangle,
\label{tstt}
\end{eqnarray} 
again in agreement with (\ref{moe}) and \cite{more}
\footnote{See previous footnote.}.
Thus, for a massless and conformally coupled scalar field, one also 
has a vanishing contribution.

\section{The Back Reaction on the Metric.}

In this Section, we shall  discuss the back reaction on the BTZ ground 
state due to the quantum fluctuations. Since any temperature is
admissible,
we choose $\beta=\infty$, which corresponds to fix the temperature of the
ground state at the lowest possible value $T=0$. \\ 
To begin with, we rewrite the 
semiclassic Einstein equations in the form ($\Lambda=-1$)
\begin{eqnarray}
{\mathcal{R}}_{\mu \nu} & = & 
-2g_{\mu \nu}+8\pi G \left( \langle T_{\mu \nu}\rangle- 
g_{\mu \nu}\langle T\rangle  
\right)\nonumber\\ 
& = & -2g_{\mu \nu}+8\pi G \langle\widehat{T}_{\mu \nu}\rangle,
\label{semi1}
\end{eqnarray}  
where we have used the result
\begin{eqnarray}
{\mathcal{R}}=-6-16\pi G \langle T\rangle.
\label{semi2}
\end{eqnarray}
Now, we have found the general expressions of the expectation 
values $\langle T_{\mu \nu}\rangle$. As a consequence, 
the semiclassic metric shows a non constant scalar curvature as well as 
a non constant Ricci tensor. Furthermore, these non constant quantities
are 
singular in the limit $r \to 0$. In the conformally coupled case, 
$\langle T\rangle$ is vanishing, but $\mathcal{R}_{\mu \nu}$ is still not 
constant and eventually one has to deal with a ``distorted'' black hole 
solution, whose nature comes from solving the semiclassical back reaction 
equations at first order in the Plank length $G$. To this aim, it is an 
usual approach starting from the general static radial symmetric solution 
in the coordinates $(t,r,\varphi)$, the ones of our background, namely 
the ground state of the BTZ solution. 
Now a  subtle point arises: in this background the one-loop approximation
may 
break down (fluctuations in $\langle T_{\mu\nu}\rangle$ would 
be of the same order of $\langle T_{\mu\nu}\rangle$). In order to cure
this
flaw, a possible trick consists in considering $N$ independent 
scalar fields instead of one, 
with  $N$ very large such that $NG=\widehat{G}$ is small and fixed 
\cite{toumb,strom,lifs94}. This has two effects: from one side
the ratio of the fluctuations to $\langle T_{\mu\nu}\rangle$ becomes
negligible in proximity of the horizon; on the other side, 
the one-loop approximation may become almost exact, because higher 
loop terms are of the order  $O(1/N)$. Within this new scheme of 
approximation, a quite natural ansaz which is consistent with the gauge 
of the background is \cite{york86,shir94}
\begin{eqnarray}
ds^2=-e^{2\widehat{G}\psi(r)}\left( r^2+\widehat{G}
\varepsilon(r)\right)dt^2
+\frac{1}{r^2+\widehat{G} \varepsilon(r)}dr^2+r^2d\varphi^2.
\label{rss}
\end{eqnarray}     
Denoting
\begin{eqnarray}
A(r)=\left( r^2+\widehat{G} \varepsilon(r)\right)^{-1}, \quad
B(r)=e^{2\widehat G\psi(r)}A^{-1}(r)
\end{eqnarray}
a standard calculation leads to
\begin{eqnarray}
R_0^0 & = & -\frac{B''}{2AB}+\frac{B'}{4AB}
\left(\frac{A'}{A}+\frac{B'}{B} \right)-\frac{B'}{2rAB},
\label{scal01} \\ 
R_1^1 & = & -\frac{B''}{2AB}+\frac{B'}{4AB}
\left(\frac{A'}{A}+\frac{B'}{B} \right)+\frac{A'}{2rA^2},
\label{scal1} \\ 
R_2^2 & = & \frac1{2rA}\left(\frac{A'}{A}-\frac{B'}{B}\right).
\label{scal0}
\end{eqnarray}
The Einstein equation associated with the mixed $(0,0)$ components gives
\begin{eqnarray}
\varepsilon'(r)=16\pi r\langle T^0_0(r)\rangle,
\label{sdfg}
\end{eqnarray}
and a suitable combination of these components leads also to
\begin{eqnarray}
-r\psi'(r)=8\pi\left(\langle T^0_0(r)\rangle-\langle T^1_1(r)\rangle
\right)
+O(\widehat{G}),  
\label{sd1o}
\end{eqnarray}
where, in the second equation, we have retained only the leading term in
$\widehat{G}$.
As solutions of the two differential equations above, we may take
\begin{eqnarray}
\varepsilon(r)&=&16\pi \int dr\, r \langle T^0_0(r)\rangle, \\ 
\psi(r)&=&8\pi \int \frac{dr}{r}\left(  \langle T^1_1(r)\rangle-\langle
T^0_0(r)\rangle \right),
\label{s1}
\end{eqnarray}
the constants of integration chosen in order to have the ground state
($M=0$) 
solution when the back reaction is switched off.
In the conformally coupled case, the computation is easier and, within our
choice of the integration constants, one has
\begin{eqnarray}
\varepsilon\left(r\right) & = & -\frac{\zeta_R\left(3\right)}{\pi^3r}+\Phi
\left(r\right), \label{solo1} \\ 
\psi\left(r\right) & = & -\frac1{4\ell}\sum_{n=1}^\infty\left(1+
\frac{\pi^2n^2r^2}{\ell^2}\right)^{-\frac32}, \label{solo2}
\end{eqnarray}
where
\begin{eqnarray}
\Phi\left(r\right)=\frac1{2\pi^2}\sum_{n=1}^\infty\frac1{n^2}\left[
\pi^2n^2r^2\left(1+\pi^2n^2r^2\right)^{-\frac32}
+2\left(1+\pi^2n^2r^2\right)^{-\frac12}\right].
\end{eqnarray}
Notice that the two series $\psi(r)$ and $\Phi(r)$ converge as long as
$r>0$.\\ 
As anticipated, a curvature singularity is present at $r=0$, but this 
singularity may be hidden by the quantum corrections as soon as there
exist 
positive real solutions to the equation
$g^{11}=0$, \emph{i.e}
\begin{eqnarray}
\widehat G\Phi\left(r\right)=\frac{\widehat
G\zeta_R\left(3\right)}{\pi^3r}-
r^2. 
\label{ecua}
\end{eqnarray}
Let us consider this equation for $r>0$.\\ 
$\Phi(r)$ is a smooth, monotonically non-increasing, and strictly positive 
function of $r$ with a unique flex at $r= r_f$ near $r=0$; 
moreover it takes the limit $\zeta_R(2)/\pi^2$ for $r\to0^+$ and
vanishes for $r\to +\infty$.
On the other hand, the function which appears in the right hand side of
(\ref{ecua}), is 
smooth and monotonically non-increasing too; furthermore, it is
positive for $r^3<\frac{\widehat G\zeta_R(3)}{\pi^3}$, divergent in the
limit 
$r\to0^+$, and shows the unique
flex in $r^3=\frac{\widehat G\zeta_R(3)}{\pi^3}$ where the function takes
the only zero in the considered domain.\\ 
In this way, it remains proved that, for each values of $\widehat G$
there exists at least one and at most three 
positive and real solutions to the equation (\ref{ecua}), 
so that the singularity $r=0$ is always shielded by an event horizon, the 
radius of which coincides with the rightmost zero where $g^{11}$ changes
sign 
(such a zero always exists); notice that, after that zero, $g^{11}>0$. 
In any cases, when $\widehat G$ is small sufficiently 
($\widehat G <\pi^3 r_f^3/\zeta_R(3)$), only one zero arises where
$g^{11}$ changes sign. Restoring the correct physical dimensions,
the event horizon satisfies
\begin{eqnarray}
0<r_+<\left[\frac{\widehat G\ell^2\zeta(3)}{\pi^3}\right]^{\frac13}
\end{eqnarray}
which, anyhow, cannot be arbitrarily large.
Qualitatively, we expect the non-conformally coupled case to be similar to
the 
one discussed here. Furthermore, the singularity dressing phenomenon 
illustrated here for the massless BTZ black hole has a four dimensional 
analogue \cite{card98} associate with the recent discovery of a class of
four 
dimensional AdS topological black holes \cite{mann97,vanzo97,brill97}.

\section{Concluding Remarks.}

In this paper, one loop quantum properties of the ground state of the BTZ
black hole have been discussed in detail, considering a scalar quantum
field
propagating in the classical background of the massless BTZ black hole.
No restriction to the gravitational coupling and the mass
of the scalar field has been 
assumed and the one-loop effective action and the expectation value for 
the energy-momentum stress tensor have been computed. As applications of 
these results, the leading order quantum correction to the BTZ black hole
entropy and the back reaction to the classical metric due to the quantum
fluctuations have been presented. With regard to the latter, we have
confirmed
that, in the presence of $N$ conformally coupled scalar field and in the 
large $N$ limit, the quantum fluctuations tend to dress the original naked 
singularity, similarly 
to the effect found in the four dimensional case \cite{card98}. This may
be 
interpreted as a quantum implementation of the Cosmic Censorship
Hypothesis.

\section*{Acknowledgments.}

It is a pleasure to thank Roberto Balbinot and  Marco Caldarelli for
useful 
discussions.

\section{Appendix.}

In this appendix, we shall briefly outline the computation of the 
heat-kernel trace for the scalar Laplace operator on the non-compact 
hyberbolic space ${\mathbb{H}}^3$, starting from the spectral theorem. 
Although the final result is well known (see, for example, 
\cite{camp90-196-1,byts96-266-1}), we think that it is useful to present
here 
an elementary derivation.\\ 
To begin with, let us introduce the operator $L=-\Delta-1$, $\Delta$ being
the 
Laplace operator on ${\mathbb{H}}^3$. Thus
\begin{eqnarray}
K_t^{{{\mathbb{H}}}^3}\!\!\left({\mathbf{x}},{\mathbf{x}}' \vert
\Delta\right)=
\langle {\mathbf{x}} \vert e^{t\Delta} \vert {\mathbf{x}}' \rangle=
e^{-t}\langle {\mathbf{x}} \vert e^{-tL} \vert {\mathbf{x}}' \rangle,
\label{a189}
\end{eqnarray}
where ${\mathbf{x}}=(y,{\mathbf{w}}) \in {\mathbb{H}}^3$.
Our aim is so to compute the heat-kernel 
$\langle {\mathbf{x}} \vert e^{-tL} \vert {\mathbf{x}}' \rangle$. 
The eigenvalues equation for $L$ is
\begin{eqnarray}
L\psi=\left[-y^2(\Delta_2+\partial^2_y)+y\partial_y-1\right]\psi=
\lambda^2 \psi,
\label{a222}
\end{eqnarray}
where $\Delta_2$ is the Laplace operator on ${\mathbb{R}}^2$ (the
transverse 
manifold), which satisfies the eigenvalues equation
\begin{eqnarray}
-\Delta_2 f_{{\mathbf{k}}}({\mathbf{w}})=k^2
f_{{\mathbf{k}}}({\mathbf{w}}),
\end{eqnarray}
where
\begin{eqnarray}
f_{{\mathbf{k}}}({\mathbf{w}})=
\frac{e^{i{\mathbf{k}} \cdot {\mathbf{w}}}}{2\pi}, 
\quad k^2={\mathbf{k}}\cdot{\mathbf{k}}.
\label{a333}
\end{eqnarray}
With the ansaz
\begin{eqnarray}
\psi=\phi(y) f_{{\mathbf{k}}}({\mathbf{w}}),
\label{4}
\end{eqnarray}
one gets the equation
\begin{eqnarray}
y^2 \phi''-y\phi'+(\lambda^2+1-k^2y^2)\phi=0,
\label{a55}
\end{eqnarray}
whose solutions  are  MacDonald's functions 
\begin{eqnarray}
\phi(y)=y K_{i\lambda}(ky),
\end{eqnarray} 
with $\lambda $ non negative.
As a result, the spectrum is continuous and the generalized eigenfunctions
are
\begin{eqnarray}
\psi_{\lambda,{\mathbf{k}}}({\mathbf{x}})=
y K_{i\lambda}(ky)f_{{\mathbf{k}}}({\mathbf{w}}).
\label{a66}
\end{eqnarray}
The non trivial spectral measure, which plays an important role, is given
by 
\begin{eqnarray}
\mu(\lambda)=\frac{2}{\pi^2} \lambda \sinh \pi \lambda.
\label{a77}
\end{eqnarray}
The spectral theorem yields
\begin{eqnarray}
\langle{\mathbf{x}}\vert e^{-tL}\vert {\mathbf{x}}'\rangle=
\int_0^\infty d\lambda\, 
\mu(\lambda)e^{-t\lambda^2}\int\frac{d^2k}{2\pi}
e^{i{\mathbf{k}} \cdot {\mathbf{u}}}
yy'K_{i\lambda}(ky)K_{i\lambda}(ky'),
\label{a88}
\end{eqnarray}
where ${\mathbf{u}}={\mathbf{w}}-{\mathbf{w}}'$. 
The integral over ${\mathbf{k}}$ can be done, making use of polar
coordinates 
in the plane, and gives 
\begin{eqnarray}
\int_0^\infty dk\, k \int_0^{2\pi} d\theta\, e^{i k 
u \cos \theta}yy'K_{i\lambda}(ky)K_{i\lambda}(ky')
=\frac{\lambda^2}{\mu(\lambda)}
\frac{P^{-\frac{1}{2}}_{i\lambda-\frac{1}{2}}(\cosh 
\sigma({\mathbf{x}},{\mathbf{x}}'))}
{\sqrt{2\pi \sinh \sigma({\mathbf{x}},{\mathbf{x}}')}}.
\label{a99}
\end{eqnarray}
Since
\begin{eqnarray}
P_{i\lambda-\frac12}^{-\frac12}
\left(\cosh\sigma\!\left({\mathbf{x}},{\mathbf{x}}'\right)\right)=
\sqrt{\frac2{\pi}}
\frac{\sin\lambda\sigma\!\left({\mathbf{x}},{\mathbf{x}}'\right)} 
{\sqrt{\sinh\sigma\!\left({\mathbf{x}},
{\mathbf{x}}'\right)}},
\label{a100}
\end{eqnarray}
an elementary integration over $\lambda$ gives
\begin{eqnarray}
K_t^{{{\mathbb{H}}}^3}\!\!\left({\mathbf{x}},{\mathbf{x}}' \vert L\right)=
\frac1{\left(4\pi t\right)^{\frac32}}
\frac{\sigma\!\left({\mathbf{x}},{\mathbf{x}}'\right)}
{\sinh\sigma\!\left({\mathbf{x}},{\mathbf{x}}'\right)}
\exp\left[-\frac{\sigma^2\!\left({\mathbf{x}},{\mathbf{x}}'\right)}{4t}
\right],
\label{a111}
\end{eqnarray}
from which (\ref{hhk}) easily follows.\\ 
Along the same lines, we determine the generalized eigenfunctions of 
the Laplace operator on the ground state solution ${\mathcal{H}}^3_0$. It
is 
convenient again to deal with the operator $L$. One has a continuous 
and discrete spectrum, because now, the transverse manifold is a 
compact 2-dimensional torus.
One has
\begin{eqnarray}
\psi_{\lambda,{\mathbf{0}}}(x)= y^{1+i\lambda}\,\qquad
\psi_{\lambda,{\mathbf{k}}}(x)= y 
K_{i\lambda}(ny)\frac{e^{i {\mathbf{w}} \cdot{\mathbf{n}}}}
{\sqrt{2\pi \beta}},
\label{mmm}
\end{eqnarray}
where ${\mathbf{x}}=(2\pi n_1/\beta, n_2)$.
As a result, the kernel of the operator $F(L)$, where $F(.)$ is a 
smooth function, reads
\begin{eqnarray}
\langle{\mathbf{x}}\vert F(L)\vert{\mathbf{x}}'\rangle & = &
\int_0^\infty d\lambda F(\lambda)y'^{1-i\lambda}y^{1+i\lambda}\nonumber \\ 
& & +\sum_{{\mathbf{k}} 
\neq {\mathbf{0}}} \frac{yy'}{2\pi \beta}\int_0^\infty d\lambda 
\mu(\lambda)F(\lambda)e^{i {\mathbf{n}} \cdot {\mathbf{u}}}
K_{i\lambda}(ny)K_{i\lambda}(ny')
\:.\label{fkk}
\end{eqnarray}

\end{document}